\documentclass[aps,showpacs,showkeys,preprint,nofootinbib]{revtex4}
\begin{document}
\newcommand{\be}{\begin{equation}}
\newcommand{\ee}{\end{equation}}
\bibliographystyle{unsrt}

\title{Langevin approach to the Porto system }
\author{J. Bok and V. \v{C}\'{a}pek}
\affiliation
{Institute of Physics of Charles University, Faculty
of Mathematics and Physics, Ke Karlovu 5, CZ-121 16 Prague 2,
Czech Republic \\(Tel. +420-2-21911338 and +420-2-21911330, Fax
+420-2-24922797, E-mail bok@karlov.mff.cuni.cz and
capek@karlov.mff.cuni.cz)}
\date{
\today
}
\begin{abstract} M. Porto (Phys. Rev. E 63 (2001) 030102) suggested a
system consisting of Coulomb interacting particles, forming a
linear track and a rotor, and working as a molecular motor. Newton
equations with damping for the rotor coordinate on the track $x$,
with a prescribed time-dependence of the rotor angle $\Theta$,
indicated unidirectional motion of the rotor. Here, for the same
system, the treatment was generalized to nonzero temperatures by
including stochastic forces and treating both $x$ and $\Theta$ via
two coupled Langevin equations. Numerical results are reported for
stochastic homogeneous distributions of impact events and Gaussian
distributions of stochastic forces acting on both the variables.
For specific values of parameters involved, the unidirectional
motion of the rotor along the track is confirmed, but with a
mechanism that is not necessarily the same as that one by Porto.
In an additional weak homogeneous potential field $U(x)=$const$
\cdot x$ acting against the motion, the unidirectional motion
persists. Then the rotor accumulates potential energy at the cost
of thermal stochastic forces from the bath.
\end{abstract}

 \maketitle
\newpage
\section{Introduction}
There are many constructions of so called ratchet systems devised
to turn stochastic perturbations or noise acting on classical or
quantum systems into a unidirectional motion. Perhaps the most
famous example is the `wind-mill' system analyzed by Feynman
\cite{FeLeSa}. Detailed analysis mostly revealed that such systems
could work but in experimental set-ups devised to violate the
Second law of thermodynamics, they fail \cite{FeLeSa,Muss}. This
fully corresponds to results of multiple theoretical analyses.

Surprisingly, the uni-directionality of the motion has recently been found
to be determined not only by a type of asymmetry of the (say) ratchet
potentials incorporated but rather dynamically
\cite{PoUrKl1,PoUrKl2,PoUrKl3}. The activity in this direction resulted
into construction of a relatively realistic model consisting of a
molecular rotor on a linear track, both consisting of particles
interacting by just Coulomb forces \cite{Port}. Under a prescribed
time-dependence of the rotor angle $\Theta(t)$ entering the potential energy of
the system, the classical Newton equation for its linear coordinate $x$
along the track yields that the rotor unidirectionally moves along the
track. The prescribed time-dependence $\Theta(t)$ may also be stochastic
what introduces the idea that one can easily convert the system, upon
adding a potential $U(x)$ slightly increasing in the direction of the rotor
motion, into a machine converting energy of the stochastic influence of the
bath, that might be of thermal origin, into a mechanical potential one.
This would mean so called perpetuum mobile of the second kind and a
violation of the Second law of thermodynamics.

Porto explicitly refuses this possibility as far as his
calculations were concerned - see a note in this respect on page 3
of \cite{Port}. Our more general calculation below, however,
confirms the above conjecture. Here, we should like to add that this
fact is by no means objection against the Porto system and its
analysis using the classical Newton equations. Instead, our
treatment rather generalizes that one of \cite{Port}. First, one
can easily justify adding stochastic forces to the Newton
equations, converting them into the Langevin ones. In this
respect, our calculations below provide a proper extension of
those by Porto \cite{Port} to finite temperatures. The Langevin
equations have well understood quantum counterparts in the Mori or
Tokuyama-Mori identities \cite{Mori,TokMor,FicSau} what makes the
calculations even more relevant. Second, the rotor angle $\Theta$
has not been here ascribed any specific time-dependence like in
\cite{Port} but has been assumed to change according to another
Langevin equation with the same temperature entering the
stochastic force correlations functions; this is rather more
physical. In fact, one should not be very much surprised by the
challenge to the Second law provided by this amended treatment of
the Porto model generalized to finite temperatures. The reason is
twofold:
\begin{itemize} \item First, there is at present increasing evidence, both
theoretical \cite{Cape1,Cape2,CapBok,Niku,Cape3,CapMan} and
experimental \cite{Shee,NikZhi,DuKuNi} as well as combined one
\cite{CapShe}, in favor of potential violability of the Second
law. \item Second, the present model and its present treatment,
though they are perhaps more realistic than often in similar
situations, are still just theoretical and their relation to
Nature may be not as obvious as it might seem at the first sight.
\end{itemize}

\section{Formulation of the problem}
In contrast to the Porto study, we impose no prescribed
time-dependence of the $\Theta(t)$ variable but describe both the
linear coordinate $x(t)$ and the $\Theta(t)$ angle as two
generalized coordinates describing the rotor (its position as well
as the rotation) with their time-dependence determined by a pair
of the corresponding coupled Langevin equations. This method
supersedes and definitely surmounts the Newton equations used by
Porto, in at least the sense that \begin{itemize} \item the
$\Theta(t)$ variable has its time-dependence determined also from
a dynamic equation, and
\item this approach respects existing connections between
dissipation (friction) incorporated and by its effect decisive in the Porto
model, and properties of the stochastic forces on the right hand side of
the Langevin equations that were completely ignored in \cite{Port}.
\end{itemize}
So, the basic set of equations reads
\be m\ddot{x}+\eta\dot{x}=-\frac{\partial}{\partial x}\Phi(x,\Theta)+
\Gamma_x(t), \quad J\ddot{\Theta}+\kappa\dot{\Theta}=
-\frac{\partial}{\partial\Theta}\Phi(x,\Theta)+\Gamma_{\Theta}(t).
\label{Langev} \ee
Here $m$, $J$, $\eta$ and $\kappa$ are the rotor mass, its moment of
inertia, and the linear and angle friction constants. As for the
stochastic forces $\Gamma_x(t)$ and $\Gamma_{\Theta}(t)$, they should
fulfil standard equations
\be \langle\Gamma_x(t)\Gamma_x(t')\rangle=2\eta
k_BT_K\delta(t-t'),  \quad
\langle\Gamma_{\Theta}(t)\Gamma_{\Theta}(t')\rangle=2\kappa
k_BT_K\delta(t-t'), \quad
\langle\Gamma_x(t)\Gamma_{\Theta}(t')\rangle=0 \label{correl} \ee
ensuring, in absence of the potential $\Phi(x,\Theta)$, that the
asymptotic mean squared velocities $\langle(\dot{x})^2\rangle$ and
$\langle(\dot{\Theta})^2\rangle$ fulfil the equipartition theorem
with temperature $T_K$. Here $\langle\ldots\rangle$ designates the
ensemble average. The stochastic forces are here represented as
\be \Gamma_x(t)=\sum_if_i^x\delta(t-t_i^x), \quad
\Gamma_{\Theta}(t)=\sum_if_i^{\Theta}\delta(t-t_i^{\Theta}), \label{Gamrep}
\ee
where $t_i^x$ and $t_i^{\Theta}$ are statistically independent times of
impact events. Times between two such succeeding impacts are exponentially distributed
\be w^x(t)=\frac{1}{\bar{t_x}}\exp(-t/\bar{t_x}), \quad
w^{\Theta}(t)=\frac{1}{\bar{t_{\Theta}}}\exp(-t/\bar{t_{\Theta}})
\label{xTheDis} \ee
($\bar{t_x}$ and $\bar{t_{\Theta}}$ being the mean waiting times) while
distributions $w^{f^x}(f)$ and $w^{f^{\Theta}}(f)$ of (also stochastic and
statistically independent) impact `forces' $f_i^x$ and $f_i^{\Theta}$ are,
on grounds of physical arguments, assumed Gaussian. For (\ref{correl}) to be
satisfied, these distributions should read
\be w^{f^x}(f)= \sqrt{\frac{1}{4\pi\eta\bar{t_x}k_BT_K}}{\rm e}^{-f^2/
(4\eta\bar{t_x}k_BT_K)}, \quad
w^{f^{\Theta}}(f)= \sqrt{\frac{1}{4\pi\kappa\bar{t_{\Theta}}k_BT_K}}{\rm
e}^{-f^2/ (4\kappa\bar{t_{\Theta}}k_BT_K)}. \label{fordis} \ee

In order to get the model fully defined, on must specify the
potential energy $\Phi(x,\Theta)$. It is connected with the
definition of the system as given by Porto \cite{Port}. The track
consists of charges $q>0$ at ${\bf
t}_n^{(+)}\equiv\{(0.1+n)b,0,-0.25b\}$ and $-q$ at ${\bf
t}_n^{(-)}\equiv\{(0.4+n)b,0,-0.25b\}$ with $n$ being arbitrary
integer. So, the track is neutral, periodic but without inversion
symmetry. The rotor has the total mass $m$ and apart from
potentially other neutral atoms, it consists of four point charges
$q_{\mu,\nu}=(-1)^{1+\mu}q'$, $\mu,\,\nu=1$ or $2$. Designating as
above position of the rotor center of mass as $x$, their positions
are ${\bf x}_{\mu,\nu}=\{x,0,0\}+r_{\mu}\{
\sin(\Theta+\Delta\Theta_{\mu,\nu}),0,\cos(\Theta+\Delta\Theta_{\mu,
\nu})\}$, $\Delta\Theta_{\mu,\nu}\equiv(\nu-\mu/2)\pi$. The two
radii are $r_1\equiv[0.005+0.045{\rm Max}(-\gamma,0)]$ and
$r_2\equiv[0.005+0.045{\rm Max}(\gamma,0)]$. The parameter
$\gamma$, $-1\le\gamma\le+1$, determines the `gear' of the Porto
molecular motor model. With that, the potential energy
$\Phi(x,\Theta)$ of the configuration of the charges reads
\be \Phi(x,\Theta)=qq'\sum_{\mu,\nu=1}^2(-1)^{1+\mu}\sum_{n=-\infty}^{
+\infty}\left[\frac{1}{|{\bf t}_n^{(+)}-{\bf x}_{\mu,\nu}(x,\Theta)|}-
\frac{1}{|{\bf t}_n^{(-)}-{\bf x}_{\mu,\nu}(x,\Theta)|}\right].
\label{potene} \ee
Calculation of the potential $\Phi(x,\Theta)$ requires some care because
the individual sums $\sum_{n=-\infty}^{+\infty}\frac{1}{|{\bf
t}_n^{(\pm)}-{\bf x}_{\mu,\nu}(x,\Theta)|}$ are divergent.

\section{Numerical treatment}
Before numerical treatment, it is useful to reformulate the problem in
dimensional units. We introduce dimensionless coordinates and time by
\be x(t)=X(T)\cdot b,\quad t=\bar{t_x}\cdot T. \label{dimleu} \ee
Then (\ref{Langev}) reads
\[ \frac{d^2}{dT^2}X(T)+{\cal E}\frac{d}{dT}X(T)=-{\cal
F}_X\frac{\partial}{\partial X}\varphi(X,\Theta)+\gamma_X(T), \] \be
\frac{d^2}{dT^2}\Theta(T)+{\cal K}\frac{d}{dT}\Theta(T)=-{\cal F}_{
\Theta}\frac{\partial}{\partial\Theta}\varphi(X,\Theta)+\gamma_{\Theta}(T)
\label{Langev2} \ee
where
\[ {\cal E}=\frac{\eta}{m}\bar{t_x}, \quad {\cal K}=\frac{\kappa}{J}
\bar{t_x}, \quad {\cal F}_X=\frac{qq'(\bar{t_x})^2}{mb^3}, \quad
{\cal F}_{\Theta}=\frac{qq'(\bar{t_x})^2}{Jb},\]
\be \varphi(X,\Theta)=\sum_{\mu,\nu=1}^2(-1)^{1+\mu}\sum_{n=-\infty}^{
+\infty}\left[\frac{b}{|{\bf t}_n^{(+)}-{\bf x}_{\mu,\nu}(Xb,\Theta)|}-
\frac{b}{|{\bf t}_n^{(-)}-{\bf x}_{\mu,\nu}(Xb,\Theta)|}\right].
\label{potene2} \ee
Formulae (\ref{Gamrep}) are then replaced by
\be \gamma_X(T)=\sum_iF_i^X\delta(T-T_i^X), \quad
\gamma_{\Theta}(T)=\sum_iF_i^{\Theta}\delta(T-T_i^{\Theta})
\label{gamrep} \ee
where, instead of (\ref{xTheDis}), we have distributions of the
(statistically independent) dimensionless times $T_i^X$ and $T_i^{\Theta}$
\be W^X(T)=\exp(-T), \quad
W^{\Theta}(T)=\frac{\bar{t}_x}{\bar{t^{\Theta}}}\exp(-T\frac{\bar{t}_x}
{\bar{t^{\Theta}}}). \label{XTheDis} \ee
As for the distributions of the (again statistically independent) `forces'
$F_i^X$ and $F_i^{\Theta}$, we have
\[ W^{F^X}(F)=\sqrt{\frac{\beta m^2b^2}{4\pi\eta(\bar{t_x})^3}}
\exp\left[-F^2\frac{\beta m^2b^2}{4\eta(\bar{t_x})^3}\right], \]
\be W^{F^{\Theta}}(F)=\sqrt{
\frac{\beta J^2}{4\pi\kappa\bar{t_{\Theta}}(\bar{t_x})^2}}
\exp\left[-F^2\frac{\beta J^2}{4\kappa\bar{t_\Theta}(\bar{t_x})^2}\right],
\quad \beta=1/(k_BT_K).
\label{Fordis} \ee
With that, the problem has now been solved numerically, using a random
number generator for simulating the stochasticity involved. The input data used were
\[{\cal E}=10, \quad {\cal K}=0.001, \quad
{\cal F}_X=13.79, \quad {\cal F}_{\Theta}=0.1092,\]
\be \frac{\beta mb^2}{4{\cal E}(\bar{t_x})^2}=0.0101,
\frac{\beta J}{4{\cal K}\bar{t}_x\bar{t_{\Theta}}}=127.5 \label{InNumDa} \ee
what could correspond to, e.g., $m=100\times$ the proton mass, $b=10^{-6}$
cm, $\bar{t_x}=10^{-10}$ sec, $\bar{t_{\Theta}}=10^{-8}$ sec, temperature
$T_K=300\,K$ and $J=10mb^2$.

\section{Numerical solution}
Series for $\varphi(X,\Theta)$ in (\ref{potene2}) as well as those
for its  partial derivatives converge very slowly. For numerical
solution, however, it is necessary to calculate as accurately as
possible. Using extrapolation methods, we were able to calculate
all these functions reasonably fast with precision of 12-13 digits
at least. Rotor-track potential energy $\varphi(X,\Theta)$ is a
periodic function in both variables, with period $1$ in $X$ ($b$
in $x$) and $\pi$ in $\Theta$. Form of this function for the
`forward gear' $\gamma=+1$ is shown in Fig. 1. Form of the
potential profile for $\gamma=-1$ (`reverse gear') can be deduced
from Fig. 1 using identity $\varphi(X,\Theta,\gamma=-1)=
-\varphi(X,\Theta+\pi/2,\gamma=+1)$. The respective positions of
extremes for $\gamma=+1$ and their values are $X=.1179500143638$,
$\Theta=.175365440363$, $\varphi=-.33522791779058053399124$
(minimum) and $X=.3802499856363$, $\Theta=-.175365440363$,
$\varphi=.33522791779058053399124$ (maximum). For $\gamma=-1$, the
values are $X=.3802499856363$, $\Theta=1.395430886432$,
$\varphi=-.33522791779058053399124$ (minimum) and
$X=.1179500143638$, $\Theta=1.746161767158$, $\varphi=
.33522791779058053399124$ (maximum; all the digits being valid).

Lengths of waiting times between two succeeding impact events were generated from uniformly
distributed random numbers $r$ and $r'\in(0,1)$ using relations $W^X(T)=-\ln(r)$ and
$W^{\Theta}(T)=-\frac{\bar{t}_x}{\bar{t^{\Theta}}}\ln(r')$. The Gaussian distributed impact
forces $F_i^X$ and $F_i^{\Theta}$ were generated using procedure gasdev from \cite{NumRec}.

Between any two impact events, set of equations (\ref{Langev2})
was solved using the Bulirsch-Stoer method (procedure bsstep from
\cite{NumRec}), with numerical constants as in (\ref{InNumDa}).
Values of constants $\frac{\beta m^2b^2}{4\eta(\bar{t_x})^3}$ and
$\frac{\beta J^2}{4\kappa\bar{t_\Theta}(\bar{t_x})^2}$ are then
$.0101$ and $127.5$, respectively. The ratio
$\frac{\bar{t_x}}{\bar{t_{\Theta}}}=0.01$. Initial values of $X$
and $\Theta$ were always chosen at minimum points of the potential
energy $\varphi$ (see above); initial values of $\frac{dX}{dT}$
and $\frac{d\Theta}{dT}$ were always chosen as zero.

Because of the nonlinearity of (\ref{Langev2}), one must expect a
chaotic behavior (extreme sensitivity to numerical values of
initial conditions as well as finite accuracy of the numerical
calculations). Fig.'s 2a and 2b illustrate the onset of the chaos. For our
accuracy $10^{-13}$ in integration of the differential equations
(\ref{Langev2}) and the above accuracy of calculation of
derivatives of $\varphi$, the chaos starts to appear always at
about $T=1200$.

\section{Results}
Fig.'s 3 and 4 illustrate time-dependence of solution for $X(T)$ for different sequences
of the random numbers (i.e. impact events as well as the values of the impact
forces) involved. In agreement with Porto's results, the tendency of behavior of $X(T)$
is to increase (for the `forward gear') and to decrease (for the `reverse gear') with
increasing $T$. One of the curves in Fig. 3 in interval (1,1000) provides an exception
from the rule. Nevertheless, extending the calculations up to $T=4000$, general tendency
of final increase was anyhow finally found to appear also here. Because of lack of
the random forces in the Porto calculations, no such exceptional behavior could be found
by him.

Fig. 5 shows a new remarkable feature of our calculations. As
Porto, in \cite{Port}, has in fact solved the problem for just $X$
with a prescribed dependence of $\Theta(T)$, and because his
calculations involved no stochasticity, his only mechanism for the
forward and backward motion of the rotor along the track was that
one owing to the tendency of $X(T)$ to relax fast, at any time, to
a minimum of the potential $\varphi(X,\Theta)$ with the
prescribed instantaneous value of $\Theta$. Thus, in his case,
coordinate $X(T)$ continually moves along valleys of the potential
profile, being driven by the prescribed time dependence of $\Theta$.
Because we have in our case also the stochastic impacts on the
coordinate $X$ included, the rotor motion along the track could
also be owing to the impact-induced hops of the rotor between two
neighboring minima of $\varphi$, without any essential change of
the `slow' variable $\Theta$. (In the whole interval (0,25) in
Fig. 5, $\Theta$ changes by just about $1/27$ of a turn.)
Fortunately, as Fig. 5 shows, direction of the prevailing motion
owing to this new mechanism is the same as that one owing to the
$\Theta$-induced motion of $X(T)$ by Porto. Thus, the two
mechanisms of the rotor motion along the track rather support each
other.

The whole beauty of the behavior of the system just reported
becomes explicit when we realize that the unidirectional tendency
of the rotor motion could survive even when we add a smooth
potential acting against the above two mechanisms. In order to
make the the situation more clear, we have added, to the potential
$\varphi(X,\Theta)$ above, a linear potential \be
\delta\varphi(X,\Theta)=c\cdot X, \quad c=0.01. \label{AddPot} \ee
In Fig. 6, we can see some results of the simulation. The curves
correspond to otherwise exactly the same parameters as in Fig. 3.
Worth noticing is that even the curve, that virtually corresponded
to the backwards motion of the rotor in the limited time-interval
investigated, turns now up to describe, like all other curves,
motion of the rotor against the homogeneous potential field
(\ref{AddPot}). Thus, because of the periodicity of
$\varphi(X,\Theta)$, the system starts now to accumulate the
potential energy $\delta\varphi(X,\Theta)$. The question is what
is the source of this acquired potential energy that could be
later used in any prescribed way. We can see its only source -
this is the above stochastic behavior simulating influence of a
real bath on the system. If so, the behavior of course becomes
incompatible with the standard thermodynamics. This illustrates
how the physics behind, in particular, the Second law of
thermodynamics is still little understood.

\section{Conclusions}
The generalized Porto model introduced above, as defined by the
Porto system \cite{Port} and the Langevin equations (\ref{Langev})
describing its behavior in external stochastic fields modelling
influence of the thermal bath, provides a relatively realistic
model that makes numerical studies of its time development
reasonable. Though the description by the Langevin equations is
classical, it should be taken seriously because of existing
quantum counterparts of these equations. This makes the basic
conclusions obtained relevant. These are, in particular:
\begin{itemize} \item Confirmation, at finite temperatures, that
the rotor has a tendency to move along the track in prevailingly
one direction only. This tendency was already found, in his
simplified treatment corresponding {\em inter alia} to zero
temperature, by Porto \cite{Port}.
\item Identification of another mechanism (in addition to that one found
by Porto himself) also leading the unidirectional motion of the
rotor along the train.
\item Survival of the behavior even when the motion goes against
a weak potential field. \end{itemize} The latter characteristic
behavior indicates possible incompatibility of the model and its
description via the Langevin equations with principles of the
statistical thermodynamics.

\section{Acknowledgement}
This work is a part of the research program MSM113200002 financed
by the Ministry of Education of the Czech republic.

 \newpage

\section{Figure captions}
\begin{description} \item[Fig. 1:] Plot of the dimensionless rotor-track
potential energy $\varphi$ for $\gamma=+1$ as a function of $X$
and $\Theta$.

\item[Fig. 2a:] Time dependence of angle $\Theta$ as a function of $T$ for $\gamma=+1$ and
the same values of all the input parameters as specified in the main text. In all three cases, the impact times as well as
the random forces were identical. Only in time intervals between any two individual impacts,
different integration time steps ($1/32$, $1/16$, $1/10$, $1/8$ and $1/7$ for curves a) to e),
respectively) were used. The onset of chaos appears at about $T=1200$. At $T=1000$, the values
of $\Theta$ obtained still coincided to 5 digits.
\item[Fig. 2b:] The same for $X$ as a function of $T$ for
the same values of all the input parameters. At $T=1000$, the values of $X$ obtained still coincided
to 6 digits. Notice that all the curves increase with increasing time.
\item[Fig. 3:] Dependence of $X$ on $T$ for five different sequences of the random numbers
involved for the `forward gear' $\gamma=+1$.
\item[Fig. 4:] The same for the `reverse gear' $\gamma=-1$. The same sequences of random numbers
as in Fig. 3 were used.
\item[Fig. 5:] Details of time-dependence of $X(T)$ ($\gamma=+1$) for curve `c'  of
Fig. 3 before the first impact event in $\Theta$ occurs. Times $T$ of
the first six impacts in $X$ were $4.4506$, $5.0489$, $5.9216$,
$9.7589$, $9.9581$, and $10.7401$. The first impact in $\Theta$
appears at $T=216.8751$. Worth noticing is how the impacts cause,
because of the periodicity of $\varphi$, sudden changes of $X=x/b$
(on the vertical axis) by integers.
\end{description}
\end{document}